\magnification=\magstep1

%
\hsize = 6.50truein
\vsize = 8.50truein
\hoffset = 0.0truein
\voffset = 0.0truein
\lineskip = 2pt
\lineskiplimit = 2pt
\overfullrule = 0pt
\tolerance = 2000
\topskip = 0pt
\baselineskip = 18pt
\parindent = 0.4truein
\parskip = 0pt plus1pt
\def\medskip{\vskip6pt plus2pt minus2pt}
\def\bigskip{\vskip12pt plus4pt minus4pt}
\def\smallskip{\vskip3pt plus1pt minus1pt}
\centerline{\bf The Antiferromagnetic Band Structure of La$_2$CuO$_4$
Revisited}
\bigskip
\centerline{Jason K. Perry,$^{1,2}$ Jamil Tahir-Kheli,$^{1,2}$ and 
William A. Goddard III$^2$}
\centerline{\it $^1$First Principles Research, Inc.}
\centerline{\it 8391 Beverly Blvd., Suite \#171, Los Angeles, CA 90048}
\centerline{\it $^2$Materials and Molecular Simulation Center, Beckman 
Institute}
\centerline{\it California Institute of Technology, Pasadena, CA 91125}
\bigskip
\noindent
{\bf Abstract.}  Using the Becke-3-LYP functional, we have
performed band structure calculations on the high temperature superconductor
parent compound, La$_2$CuO$_4$.  Under the restricted spin formalism
($\rho_\uparrow = \rho_\downarrow$), the
R-B3LYP band structure agrees well with the standard LDA band structure.
It is metallic with a single $Cu\ x^2-y^2/O\ p_\sigma$ band
crossing the Fermi level.  Under the unrestricted spin formalism
($\rho_\uparrow \ne \rho_\downarrow$), the U-B3LYP band structure has 
a spin polarized antiferromagnetic solution with a band
gap of 2.0 eV, agreeing well with experiment.  This state is 
1.0 eV (per formula unit) {\it lower} than that calculated from the R-B3LYP.
The apparent high energy of the spin restricted state is
attributed to an overestimate of on-site Coulomb repulsion which is
corrected in the unrestricted spin calculations.
The stabilization of the total energy with spin
polarization arises primarily from the
stabilization of the $x^2-y^2$ band, such
that the character of the eigenstates at the top of the valence band in the 
antiferromagnetic state becomes a strong mixture of
$Cu\ x^2-y^2/O\ p_\sigma$ and $Cu\ z^2/O'\ p_z$.  
Since the Hohenberg-Kohn theorem requires the spin restricted and
spin unrestricted calculations to give
identical ground state energies and total spatial
densities for the exact functionals,
this large disparity in energy reflects the inadequacy of
current functionals for describing the cuprates.  This calls into 
question the use of band structures based on current restricted spin density 
functionals (including LDA) 
as a basis for single band theories of superconductivity in these materials.  
\vfill
\eject

\centerline{\bf INTRODUCTION}
\smallskip
Almost immediately following the discovery of the superconducting
cuprates ({\it e.g.} La$_{1-x}$Ba$_x$CuO$_4$ in 1986),
several research groups characterized the band structures
within the density functional (DFT) formalism.$^{1-3}$  Using the 
standard 
local density approximation (LDA), the resulting 
band structures consistently showed the Fermi level behavior of these
materials to be characterized by a single metallic 2-D band,
comprised of $Cu\ x^2-y^2/O\ p_\sigma$ hybrid orbitals (from here on called the
$x^2-y^2$ band).  All other occupied
bands were buried 0.5 eV or more below the Fermi level.  
While the basic orbital picture appeared to agree with some experimental data,
the absence of antiferromagnetic (AF) order in the band structure of the
undoped parent compound ({\it e.g.} La$_2$CuO$_4$) was cause for
concern.  In time, the inability of this band structure to explain 
an increasingly diverse range
of normal state phenomena of the doped materials
led many to conclude that Fermi liquid
theory is not applicable.  Indeed, no major theories
based solely on this conventional band structure have survived.
Still, the LDA band structure is widely viewed as a reasonable
starting point for superconductivity theories and experimental
data is routinely compared to these calculations.

In the early 1990's, a number of groups
succeeded in computing an alternative AF band structure for
undoped La$_2$CuO$_4$ within the 
DFT formalism.  Using a pseudopotential approach, Shiraishi, {\it et al.}$^4$
achieved a spin polarized solution where $\uparrow$ and $\downarrow$ spins
reside on different Cu sites of a doubled unit cell.  This opened up a band gap
of 0.6 eV between occupied and unoccupied bands (the measured gap is
2.0 eV$^5$).  Other groups attempted to correct certain
known flaws in the LDA functional.
Most notably, Svane$^6$ applied a self-interaction correction
local spin density
(SIC-LSD) approach, in which the residual Coulomb interaction an electron
improperly sees with itself is removed from the LSD functional.
Spin localization was achieved with this method and an AF
band structure was found with an indirect band gap of 1.04 eV.  Temmerman,
{\it et al.}$^7$ later found similar results using an alternative SIC-LSD 
approach.
They reported an improved band gap of 2.1 eV.
Czyzyk and Sawatzky$^8$ took yet another approach (LSDA+U), embedding a Hubbard
Hamiltonian into the Kohn-Sham LDA equation.  They also achieved an
AF state with a band gap of 1.65 eV.

A common characteristic
of all these calculations was a significant change in
orbital character near the top of the valence band as compared to the
standard LDA band structure.  All of the above authors
noted a large increase in either the 
apical oxygen ($O'$) or $Cu\ z^2$ density of states.  While these results
suggested
the single band ($x^2-y^2$) picture of LDA may not be an adequate starting 
point for the doping range of superconductivity,
it was not immediately apparent that
a more complicated band picture was consistent with the 
experimental data either.  Unresolved was the difficult question of how to 
describe the doped state of the superconductor which appears to produce
a Fermi surface in the Brillouin zone of the single unit cell.$^9$  Removing 
electrons from a rigid band structure may be appropriate with the
standard metallic state, but this procedure is less clear when starting from 
the undoped spin polarized insulating band structure in the reduced 
Brillouin zone (doubled unit cell).  Thus, the LDA band structure has remained 
the de facto standard in the field to this day.

In this work, we revisit the DFT band structure using the Becke-3-Lee-Yang-Parr
(B3LYP) functional.$^{10}$  The superiority of this hybrid functional,
which includes a contribution of the exact Hartree-Fock exchange, has
been well documented for molecular systems.  Notably, Martin and 
Illas$^{11}$ showed the utility of such hybrid functionals in 
dramatically improving the 
calculation of the coupling constant $J$ from cluster models of La$_2$CuO$_4$. 
Unfortunately, the method is still little used to determine band structures,
partly because of the expense of such computations.
We show here that spin unrestricted B3LYP (U-B3LYP) leads to an AF band 
structure
in agreement with the SIC-LSD and LSDA+U results cited above. This assuages
doubts about that work and confirms the 
$Cu\ z^2/O'\ p_z$ character at the Fermi level.  Most
importantly, this AF state is found to be 1.0 eV
per formula unit more stable than the state calculated from the
spin restricted (R-B3LYP) functional.  
Considering the Hohenberg-Kohn theorem$^{12}$ requires the
exact spin restricted and spin unrestricted functionals to yield identical
total spatial densities ($\rho_\uparrow + \rho_\downarrow$)
and total energies,
this discrepancy indicates a serious flaw in current  
functionals (B3LYP and LDA).  Therefore, use of the LDA band structure 
to justify single band models of superconductivity in the cuprates 
is highly questionable.

\smallskip
\centerline{\bf RESULTS AND DISCUSSION}
\smallskip

Calculations were performed using CRYSTAL98$^{13}$ which employs an atomic
Gaussian type orbital basis set.  For O, the standard 8-411G basis set with
a D polarization exponent of 0.65 was used.$^{14}$  
For Cu and La, the Hay and Wadt$^{15}$
effective core potentials (ECP's) were used.  These ECP's treat explicitly 
the outer core (3s and 3p for Cu, 5s and 5p for La) and valence electrons.
The basis sets used with these ECP's were modified from the original basis
sets of Hay and Wadt, since some functions are too diffuse for
calculations on crystals.  For Cu, the two diffuse S exponents were replaced
by a single exponent optimized to 0.30 from LDA calculations on La$_2$CuO$_4$.
The two Cu diffuse P exponents were replaced by a single exponent optimized
to 0.20.
The basis set was contracted to (3s3p3d) based on atomic Cu(II)
calculations.  For La, the two diffuse S exponents were replaced with a
single exponent optimized to 0.10.  The two diffuse P exponents and the diffuse
D exponent were removed without replacement.  The
basis set was contracted to (3s2p1d) based on atomic La(III) and La(II)
calculations.  Overall, the quality of the basis set is superior to that
used by Su, {\it et al.}$^{16}$ in their CRYSTAL95 Hartree-Fock (HF) study of
La$_2$CuO$_4$.  Several alternative basis sets were tested, all leading
to similar results.  The tetragonal La$_2$CuO$_4$
crystal structure was taken from Hazen.$^{17}$

Figure 1 presents the results of our LDA and R-B3LYP calculations 
with restricted spin and tetragonal symmetry.  The LDA band
structure is in excellent agreement with previous plane wave 
calculations$^{1-3}$
and there is little difference with the R-B3LYP band structure result.
As expected for restricted spin calculations, both methods produce Pauli
paramagnetic (PM) band structures.  In each, the only band crossing
the Fermi level is the highly 2-D $x^2-y^2$ band.  The next band ($z^2$) is
approximately 1 eV below the Fermi level.  

The lack of antiferromagnetic order in these band structures limits their
usefulness for analyzing properties of the undoped material.
On the other hand,
such PM band structures may be entirely appropriate for understanding the 
nature of the doped state.  Indeed, the LDA and R-B3LYP band structures seem
to be consistent with some experimental data such as the ARPES Fermi 
surface,$^9$ but they remain incompatible with many other experiments.
An analysis of the density of
states shows the nature of the intrinsic undoped hole
(lowest unoccupied states totaling 1 hole per formula unit) is 48\% 
$Cu\ x^2-y^2$ and 47\% $O\ p_\sigma$.  While we have not explicitly carried out
computations on the doped state ($x$=0.15), we can 
estimate the nature of the doped hole using a rigid band model.  The character
of this doped hole (highest occupied states totaling 0.15 electron)
is 47\% $Cu\ x^2-y^2$, 38\% $O\ p_\sigma$, 6\% $O'\ p_z$, and 5\% $Cu\ z^2$.
The relatively small amount of $O'$ and $z^2$ character in these orbitals
is in keeping with most models for superconductivity in the cuprates.
However, x-ray absorption studies (XAS) support a total $z^2$ hole
contribution of 5\% to 20\% and a similar range for 
$O'\ p_z$.$^{18}$  Details such as the ARPES pseudogap,$^{19}$ 
the anomalous background signal,$^{20}$ and
other probes of the normal state properties such as the NMR,$^{21}$ 
resistivity,$^{22}$
and neutron scattering$^{23}$ also appear to have no explanation using this 
conventional band structure.

We find that,
a different state emerges under the unrestricted spin 
(U-B3LYP) implementation of this functional.  Using an orthorhombic unit cell,
the resulting band structure from these calculations is shown in Figure 2.
The combination of a spin functional
and a doubled unit cell allows for a possible spin polarized solution.
Indeed, we find an apparent AF state with a gap of 2.0 eV.  The
band dispersion is in excellent agreement with previously published DFT
band structures for this AF state,$^{6-8}$ and the computed gap
agrees with the measured gap.$^5$  Furthermore, a U-B3LYP calculation of
the ferromagnetic (FM) state, which has a pure $\uparrow$ spin per formula
unit under tetragonal symmetry, is found to be 0.18 eV higher in energy.
This compares favorably (noting caveats)$^{16}$ with the experimental 
$J$ value of 0.13 
eV.$^{24}$ 

A principal advantage of this new U-B3LYP AF band structure is that
it follows unambiguously using a well established 
functional.  No additional empirical corrections were necessary.
Furthermore, while no comparison of the relative
stabilities of the LDA state and the SIC-LSD (or LSDA+U) state has
been previously reported, such a comparison is rather
straightforward with our calculations.
Significantly, the U-B3LYP AF ground state and FM excited state are
found to be 1.0 and 0.82 eV per La$_2$CuO$_4$ formula unit more
stable than the R-B3LYP state.  This represents a rather dramatic
failure on the part of the spin restricted functional.  Clearly the
U-B3LYP calculation leads to a superior representation of the ground
state of La$_2$CuO$_4$.

To better understand the significance of this 1.0 eV energy difference,
one must consider the theoretical foundation of DFT.
The Hohenberg-Kohn theorem$^{10}$ proves the existence of two functionals
F$_{\rm HK}[\rho]$
and
F$_{\rm HK}[\rho_\uparrow,\rho_\downarrow]$ 
where the first is a functional of the total density
$\rho=\rho_\uparrow + \rho_\downarrow$ and the second is a functional of
the two spin densities $\rho_\uparrow$ and $\rho_\downarrow$.
For a spin-independent
potential, the first functional will lead to the exact ground state
energy and total spatial density, 
$\rho=\rho_\uparrow + \rho_\downarrow$.
For spin-dependent spatial potentials, the use of the second
functional will lead to the ground state energy and spin
densities $\rho_\uparrow$ and $\rho_\downarrow$.
In addition,
F$_{\rm HK}[\rho_\uparrow,\rho_\downarrow]$ 
can be used for the special case of a spin-independent potential
where it must obtain the same ground state energy and total
density $\rho$ as
F$_{\rm HK}[\rho]$.

The Kohn-Sham
orbitals arising from energy minimization for La$_2$CuO$_4$ will
be of restricted (doubly occupied) Hartree-Fock type for
F$_{\rm HK}[\rho]$
with $\rho(r)=2\sum_{occ}|\phi_i(r)|^2$ and of unrestricted Hartree-Fock
type for 
F$_{\rm HK}[\rho_\uparrow,\rho_\downarrow]$ 
where the $\uparrow$ spin orbitals may be different from the $\downarrow$
spin orbitals.  Both minimizations should lead to exactly the same
energy and ground state total density $\rho$.
If the R-B3LYP functional were close to F$_{\rm HK}[\rho]$
and the U-B3LYP functional were close to 
F$_{\rm HK}[\rho_\uparrow,\rho_\downarrow]$,
then the energies of the two calculations for La$_2$CuO$_4$ should also
be close.

Our computed 1.0 eV energy difference leads us to conclude
that at least one of the two functionals is not
close to the exact functional. Since the U-B3LYP band structure is
an excellent description of the AF state, we are led to question the
quality of the R-B3LYP functional.

It is well known in fact that spin density
functional approximations to
F$_{\rm HK}[\rho_\uparrow,\rho_\downarrow]$ 
are superior to total density approximations to F$_{\rm HK}[\rho]$
since it is much easier to empirically design functionals to correct
for the exchange coupling of like spins when $\rho$ is separated into
$\rho_\uparrow$ and $\rho_\downarrow$.  
In the restricted spin functionals, R-B3LYP and LDA, the improper
treatment of exchange in strongly correlated systems leads to an
overestimate of on-site Coulomb repulsion.

A classic example of the
problems that occur with these functionals is the dissociation of H$_2$.
At equilibrium distances, the molecule is well described both by closed-shell
LDA and by B3LYP functionals.  However, at the dissociation limit both 
functionals lead to an energy calculated to be higher than two H atoms
due to the ionic
components (H$^+$ + H$^-$).  With current
functionals, to properly describe dissociation requires calculation of
either the triplet state or the symmetry broken unrestricted ``singlet''
spin state, either of which leads to
net formation of magnetic moments on each H atom.
{\it This difference between the spin restricted and unrestricted
functionals
should not be considered
a failure of
density functional theory.  Instead,
it is a failure of the empirical
spin restricted
functionals that have been developed so far.}

By analogy to H$_2$, the FM state of undoped La$_2$CuO$_4$ (where each 
site has a pure $\uparrow$ spin) and
the AF state (where each site has either a pure
$\uparrow$ or a pure $\downarrow$ spin) are well described by the spin
unrestricted U-B3LYP functional.
Using the spin restricted formalism where
each site is 50\% $\uparrow$ AND 50\% $\downarrow$ (R-B3LYP),
leads not only to an overestimate (1.0 eV per formula unit) of
the total energy, but also to an incorrect band structure.
In particular, the
$x^2-y^2$ band is elevated relative to the other bands as a result of the
improper on-site Coulomb repulsion associated with the 
$Cu\ x^2-y^2/O\ p_\sigma$ orbital.  Removing this repulsion through the
localization of spins in either the FM or AF states
stabilizes the $x^2-y^2$ band, resulting in 
an increase in $z^2$ character near the
top of the valence band.

As compared to the R-B3LYP state, the added stability of the U-B3LYP total 
energy by 1.0 eV is associated with an approximately 1 eV stabilization of
the $x^2-y^2$ band with respect to the other bands.  This is most noticeable
in the nature of the doped hole.
We characterize the U-B3LYP undoped hole as 56\% $Cu\ x^2-y^2$
and 38\% $O\ p_\sigma$, a picture 
not substantially different from the
R-B3LYP calculation.  The ratio of $Cu\ x^2-y^2/O\ p_\sigma$ character is
somewhat larger in the U-B3LYP state, but qualitatively both
calculations agree that
the undoped hole states are nearly purely derived from
these two orbitals.  The picture changes substantially upon
considering the nature of the doped holes.  Figure 3 presents a detail
of the density of states in the vicinity of the Fermi level for the $x$=0.15
doped state, assuming a rigid band model.  The nature of the doped hole
is characterized as 17\% $Cu\ x^2-y^2$, 40\% $O\ p_\sigma$, 21\% $O' p_z$,
and 19\% $Cu\ z^2$.  This is summarized in Table I.  The significant increase
in the $Cu\ z^2$ and $O'\ p_z$ character of the doped hole as compared to
the R-B3LYP results is comparable to that noted in previous
AF band structure calculations.$^{4,6-8}$  While doubts about the
quality of the previous band structures undermined the significance of these
findings, the cumulative weight of these results now strongly favors
the scenario where $z^2$ holes are formed upon doping.

While the U-B3LYP band structure may be a good representation of the
AF ground state of undoped La$_2$CuO$_4$, we should not lose sight of the
fact that ultimately a spin restricted (total density) functional
should be equally successful.  Indeed, a proper PM band structure
is more useful in understanding the role of doping in superconductivity.
In our view, such a band structure has yet to be achieved with DFT, leading 
many to conclude that Fermi liquid theory has failed
for these materials. The simpler answer might just be that LDA has failed.
The calculations presented here demonstrate the shortcomings of these
methods and suggest possible solutions.  We might postulate the
existence of a modified R-B3LYP functional which gives exactly the same
total density and total energy as the U-B3LYP functional.  The PM
band structure associated with this new functional may be substantially
different from the one we've calculated here. 
Elsewhere, we have already
incorporated such a correction into a simple tight binding model for the
doped superconductor.$^{25}$  
By effectively introducing a local magnetic moment at each
Cu site within a PM model of optimally doped 
La$_{1.85}$Sr$_{0.15}$CuO$_4$, we showed that the $x^2-y^2$ band is
significantly stabilized relative to the other bands.  This brings the
narrow $z^2$ band to the Fermi level.  The resulting band structure 
has a unique crossing between the 2-D-like $x^2-y^2$ band and the
1-D-like $z^2$ band.
This reflects a dramatic first-order correction to the standard
band structure.  Indeed, this model has already been used to interpret
the ARPES pseudogap and anomalous background,$^{26}$ the NMR
Cu and O relaxations and Knight shifts,$^{27}$ the Hall effect, and Josephson
tunneling$^{28}$ have already been presented with this model.
\medskip
\centerline{\bf CONCLUSIONS}
\smallskip
In summary, we have presented the results of R-B3LYP and U-B3LYP band structure
calculations on La$_2$CuO$_4$.
The R-B3LYP results are in good agreement with previous LDA calculations 
and the U-B3LYP results are in good agreement with previous SIC-LSD 
calculations 
(among others).  The large discrepancy in energy between the two states
(1.0 eV per formula unit)
is attributed to an improper overestimate of on-site Coulomb
repulsion within the spin restricted calculations.  The automatic correction of
this error within the U-B3LYP AF state leads to the stabiliation of the
$x^2-y^2$ band relative to the other occupied bands.  As a result, the
$z^2$ band is then brought to the top of the valence band and contributes
significantly to the doped hole states.  These results cast significant
doubt on the continued use of LDA band structures as the starting point
for theories of superconductivity in these materials.

We wish to acknowledge helpful discussions with Dr. Francesco Faglioni and
Dr. Eugene Heifets.  This work was partially supported by the Materials and
Process Simulation Center (MSC) at Caltech which is supported by grants
from DOE-ASCI, ARO/DURIP, ARO/MURI, 3M, Beckman Institute, Seiko-Epson,
Dow, Avery-Dennison, Kellogg, and Asahi Chemical.
\medskip
\centerline{\bf REFERENCES}
\smallskip
\item{$^1$}J. Yu, A.J. Freeman, and J.H. Xu, Phys. Rev. Lett. {\bf 58}, 1035 
(1987).

\item{$^2$}L.F. Mattheiss, Phys. Rev. Lett. {\bf 58}, 1028 (1987).

\item{$^3$}W.E. Pickett, Rev. Mod. Phys. {\bf 61}, 433 (1989).

\item{$^4$}K. Shiraishi, A. Oshiyama, N. Shima, T. Nakayama, and H. Kamimura,
Solid State Commun. {\bf 66}, 629 (1988).

\item{$^5$}J.M. Ginder, R.M. Roe, Y. Song, R. P. McCall, J.R. Gaines, 
E. Ehrenfreund, and A.J. Epstein, Phys. Rev. B {\bf 37}, 7506 (1988).

\item{$^6$}A. Svane, Phys. Rev. Lett. {\bf 68}, 1900 (1992).

\item{$^7$}W.M. Temmerman, Z.Szotek, and H. Winter, Phys. Rev. B {\bf 47}, 
11533 (1993).

\item{$^8$}M.T. Czyzyk and G.A. Sawatsky, Phys. Rev. B {\bf 49}, 14211 (1994).

\item{$^9$}A. Ino, C. Kim, T. Mizokawa, Z.-X. Shen, A. Fujimori,
M. Takaba, K. Tamasaku, H. Eisaki, and S. Uchida, J. Phys. Soc. Japan
{\bf 68}, 1496 (1999).

\item{$^{10}$}A.D. Becke, J. Chem. Phys. {\bf 98}, 5648 (1993); C. Lee, W. Yang,
and R.G. Parr, Phys. Rev. B {\bf 37}, 785 (1988).

\item{$^{11}$}R.L. Martin and F. Illas, Phys. Rev. Lett. {\bf 79}, 1539 (1997).

\item{$^{12}$}P. Hohenberg and W. Kohn, Phys. Rev. {\bf 136}, B864 (1964).

\item{$^{13}$}V.R. Saunders, R. Dovesi, C. Roetti, M. Caus\`a, N.M. Harrison,
R. Orlando, C.M. Zicovich-Wilson, {\it CRYSTAL98 User's Manual}, University
of Torino, Torino, 1998.

\item{$^{14}$}http://www.dl.ac.uk/TCS/Software/CRYSTAL.

\item{$^{15}$}P.J. Hay and W.R. Wadt, J. Chem. Phys. {\bf 82}, 299 (1985).

\item{$^{16}$}Y.-S. Su, T.A. Kaplan, S.D. Mahanti, and J.F. Harrison,
Phys. Rev. B {\bf 59}, 10521 (1999).

\item{$^{17}$}R.M. Hazen, in {\it Physical Properties of High Temperature
Superconductors II}, ed. D.M. Ginsberg (World Scientific, New Jersey; 1990),
121-198.

\item{$^{18}$}N. N\"ucker, H. Romberg, X.X. Xi, J. Fink, B. Gegenheimer,
and Z.-X. Shen, Phys. Rev. B {\bf 39}, 6619 (1989); A. Bianconi, S. Della
Longa, C. Li, M. Pompa, A. Gongiu-Castellano, D. Udron, A. M. Flank, and
P. Lagarde, Phys. Rev. B {\bf 44}, 10126 (1991); C.T. Chen, L.H. Tjeng,
J. Kwo, H.L. Kao, P. Rudolf, F. Sette, and R.M. Fleming, Phys. Rev. Lett.
{\bf 68}, 2543 (1992).

\item{$^{19}$}M.R. Norman, H. Ding, M. Randeria, J.C. Campuzano, T. Yokoya,
T. Takeuchi, T. Takahashi, T. Mochiku, K. Kadowaki, P. Guptasarma, and
D.G. Hinks, Nature {\bf 392}, 157 (1998).

\item{$^{20}$}Z.-X. Shen and D.S. Dessau, Phys. Rep. {\bf 253}, 2 (1995).

\item{$^{21}$}R.E. Walstedt, B.S. Shastry, and S.-W. Cheong, Phys. Rev.
Lett. {\bf 72}, 3610 (1994).

\item{$^{22}$}H. Takagi, B. Batlogg, H.L. Kao, J. Kwo, R.J. Cava, 
J.J. Krajewski, and W.F. Peck, Jr., Phys. Rev. Lett. {\bf 69}, 2975 (1992).

\item{$^{23}$}S.-W. Cheong, G. Aeppli, T.E. Mason, H. Mook, S.M. Hayden,
P.C. Canfield, Z. Fisk, K.N. Clausen, and J.L. Martinez, Phys. Rev. Lett.
{\bf 67}, 1791 (1991).

\item{$^{24}$}G. Aeppli, S.M. Hayden, H.A. Mook, Z. Fisk, S.-W. Cheong,
D. Rytz, J.P. Remeika, G.P. Espinosa, and A.S. Cooper, Phys. Rev. Lett.
{\bf 62}, 2052 (1989); P.E. Sulewski, P.A. Fleury, K.B. Lyons, S.-W.
Cheong, and Z. Fisk, Phys. Rev. B {\bf 41}, 225 (1990).

\item{$^{25}$}J.K. Perry and J. Tahir-Kheli, Phys. Rev. B {\bf 58}, 12323
(1998); J.K. Perry, J. Phys. Chem. A {\bf 104}, 2438 (2000); J.K. Perry
and J. Tahir-Kheli, unpublished (cond-mat/9907332).

\item{$^{26}$}J.K. Perry and J. Tahir-Kheli, Phys. Rev. B, submitted 
(cond-mat/9908308).

\item{$^{27}$}J. Tahir-Kheli, J. Phys. Chem. A {\bf 104}, 2432 (2000).

\item{$^{28}$}J. Tahir-Kheli, Phys. Rev. B {\bf 58}, 12307 (1998).
\vfil\eject
\noindent{\bf Table I.} Orbital character of intrinsic undoped holes (totaling
1 hole), doped holes (totaling 0.15 hole), and total holes at
optimal doping (totaling 1.15 holes).  Results are shown for both
the restricted spin R-B3LYP state and unrestricted spin
U-B3LYP AF state.

\vskip 0.5truein
\halign{\noindent#\hfill &\quad \hfill#\hfill &\quad \hfill#\hfill 
&\quad \hfill#\hfill &\quad \hfill#\hfill &\quad \hfill#\hfill
&\quad \hfill#\hfill &\quad \hfill#\hfill \cr
\noalign{\bigskip\hrule\smallskip}
\noalign{\hrule\medskip}
& & B3LYP & & & & UB3LYP & \cr
Orbital & undoped & doped & total & & undoped & doped & total \cr
\noalign{\medskip\hrule\medskip}
$Cu\ x^2-y^2$ & 48\% & 47\% & 48\% & & 56\% & 17\% & 51\% \cr
$O\ p_\sigma$ & 47\% & 38\% & 46\% & & 38\% & 40\% & 38\% \cr
$O'\ p_z$ & 1\% & 6\% & 2\% & & 1\% & 21\% & 4\% \cr
$Cu\ z^2$ & 1\% & 5\% & 2\% & & 0\% & 19\% & 3\% \cr
\noalign{\medskip\hrule\smallskip}
\noalign{\hrule\bigskip}}
\vfill
\eject
\centerline{\bf FIGURE CAPTIONS}
\smallskip
\noindent{\bf Figure 1.}  Band dispersions of La$_2$CuO$_4$
plotted along symmetry lines
of the tetragonal Brillouin zone (see reference 1) from restricted spin (a) LDA
and (b) R-B3LYP calculations.  Results are in good agreement with the LDA
computations of references 1-3.

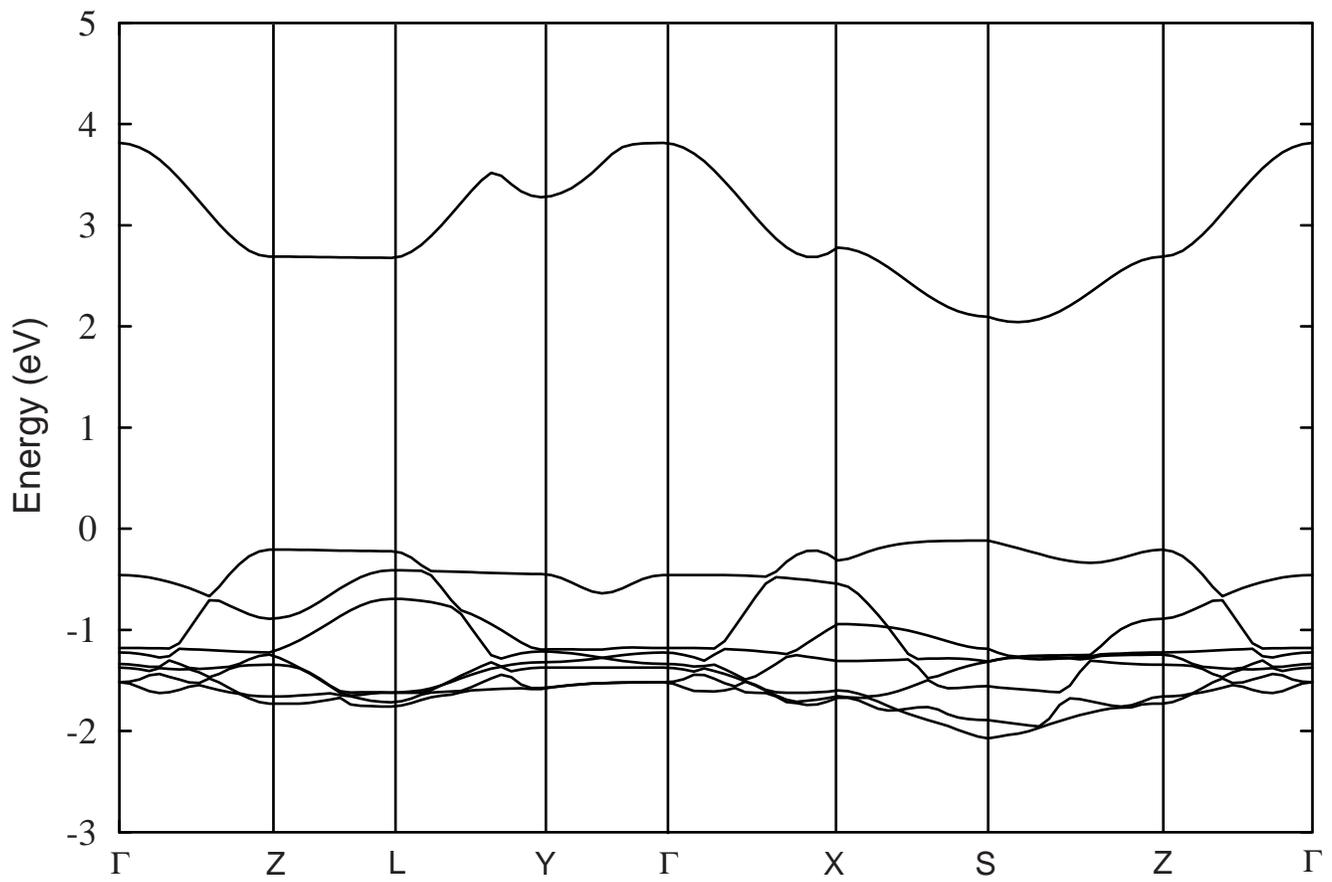
\noindent{\bf Figure 2.}  Band dispersion plotted along symmetry lines of
the orthorhombic Brillouin zone (see reference 8) from the unrestricted spin 
U-B3LYP calculation of La$_2$CuO$_4$.  
Results are in good agreement with the SIC-LSD and 
LSDA+U computations of references 6-8.

\noindent{\bf Figure 3.}  Detail of the density of states from the unrestricted 
spin U-B3LYP AF calculation of La$_{2-x}$Sr$_x$CuO$_4$ ($x=0$).  
Fermi level is positioned to a doping level of
$x$=0.15.  (a) projected Cu DOS vs. total DOS.  (b) projected O DOS vs.
total DOS.  (c) projected O' DOS vs. total DOS.  (d)  projected Cu $x^2-y^2$
DOS vs. total Cu DOS.  (e)  projected Cu $z^2$ DOS vs. total Cu DOS.
\vfill\eject
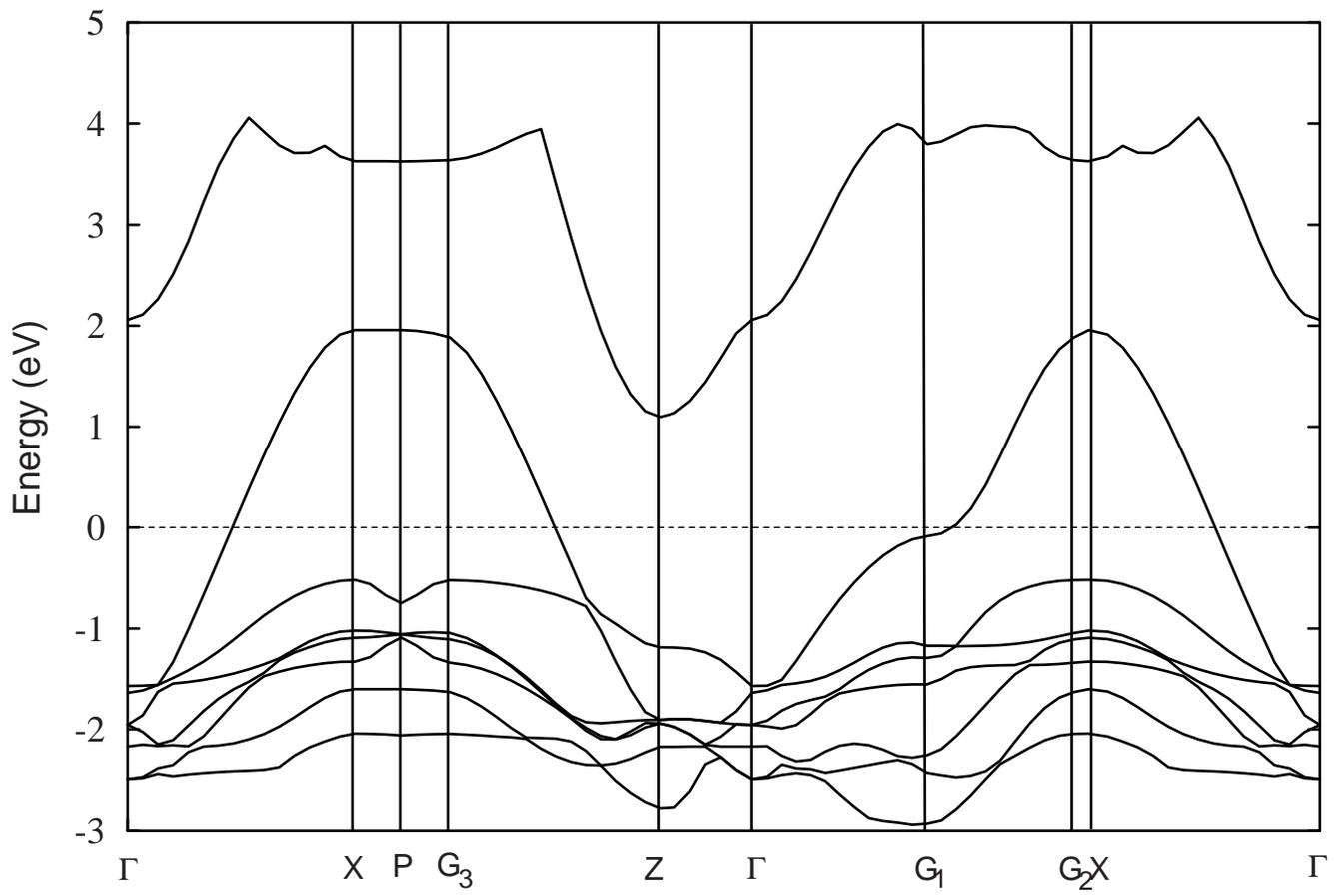
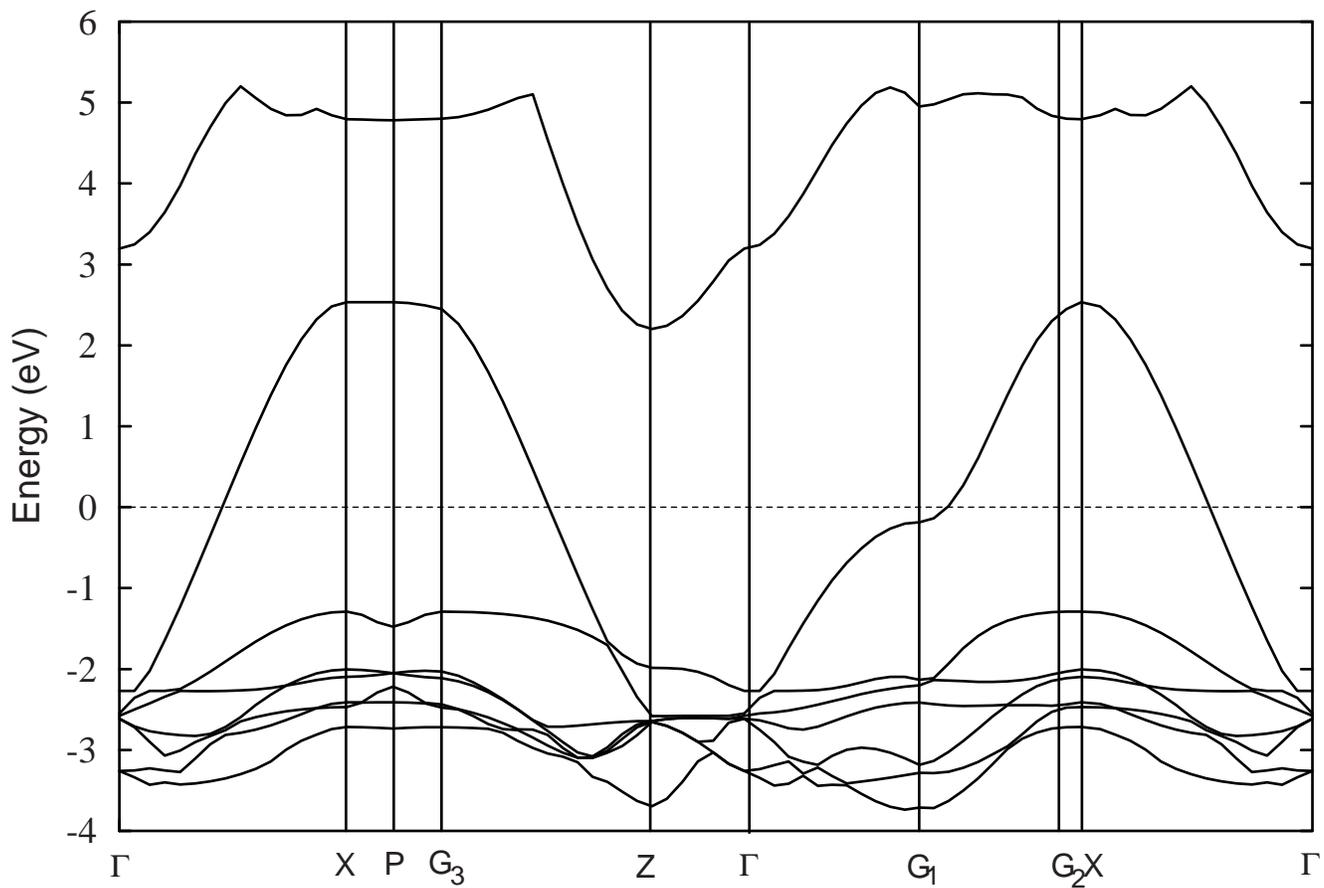
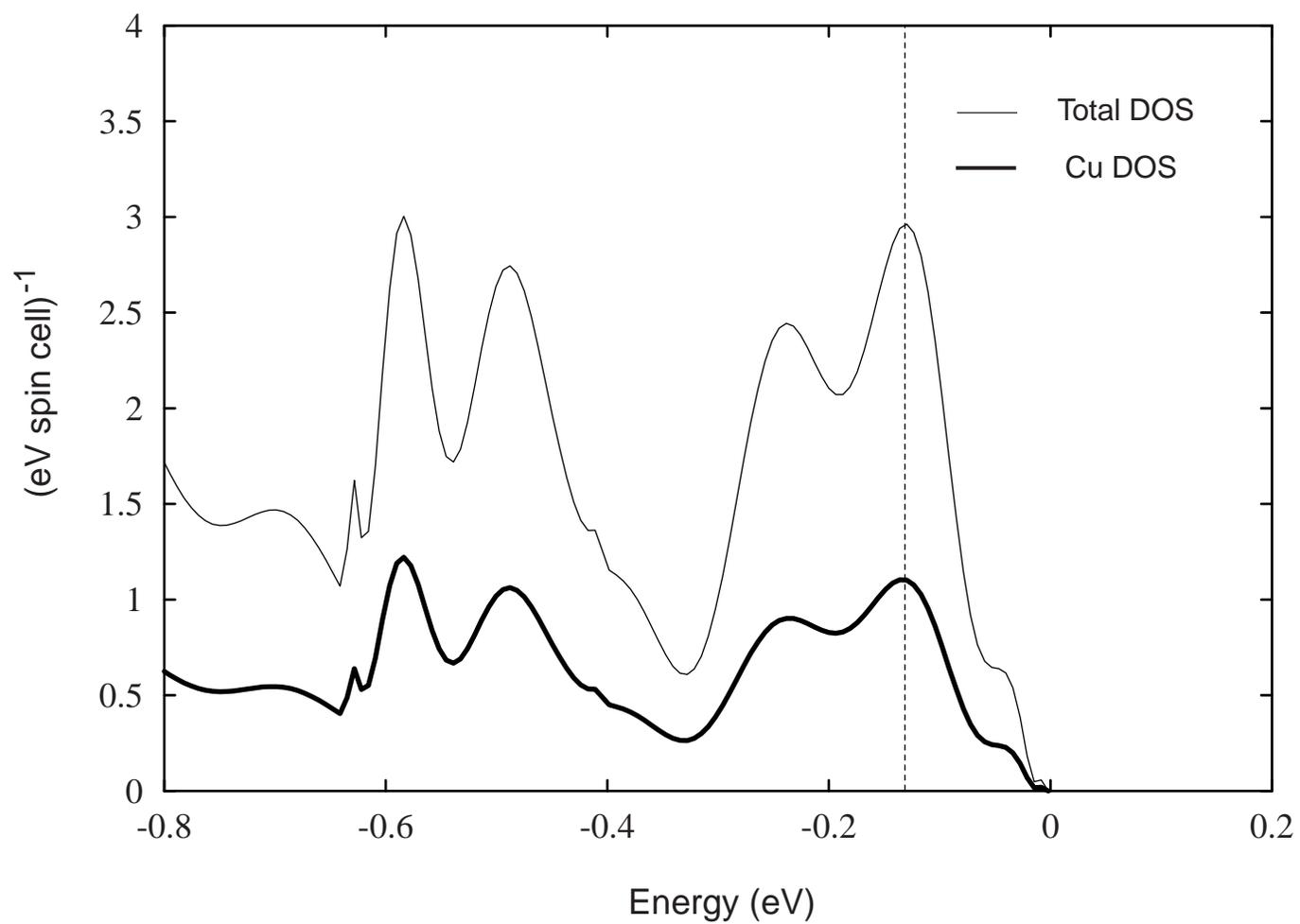
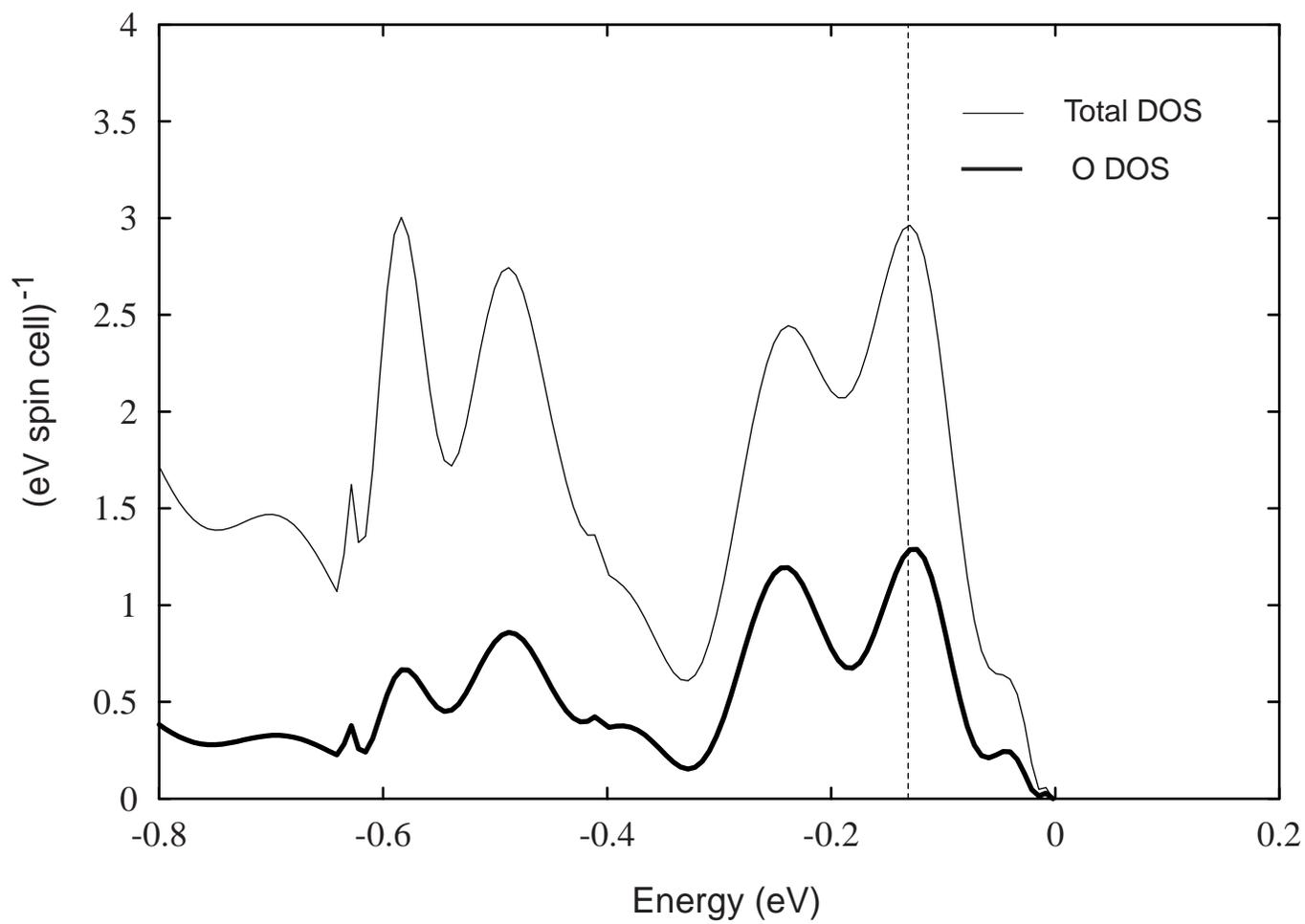
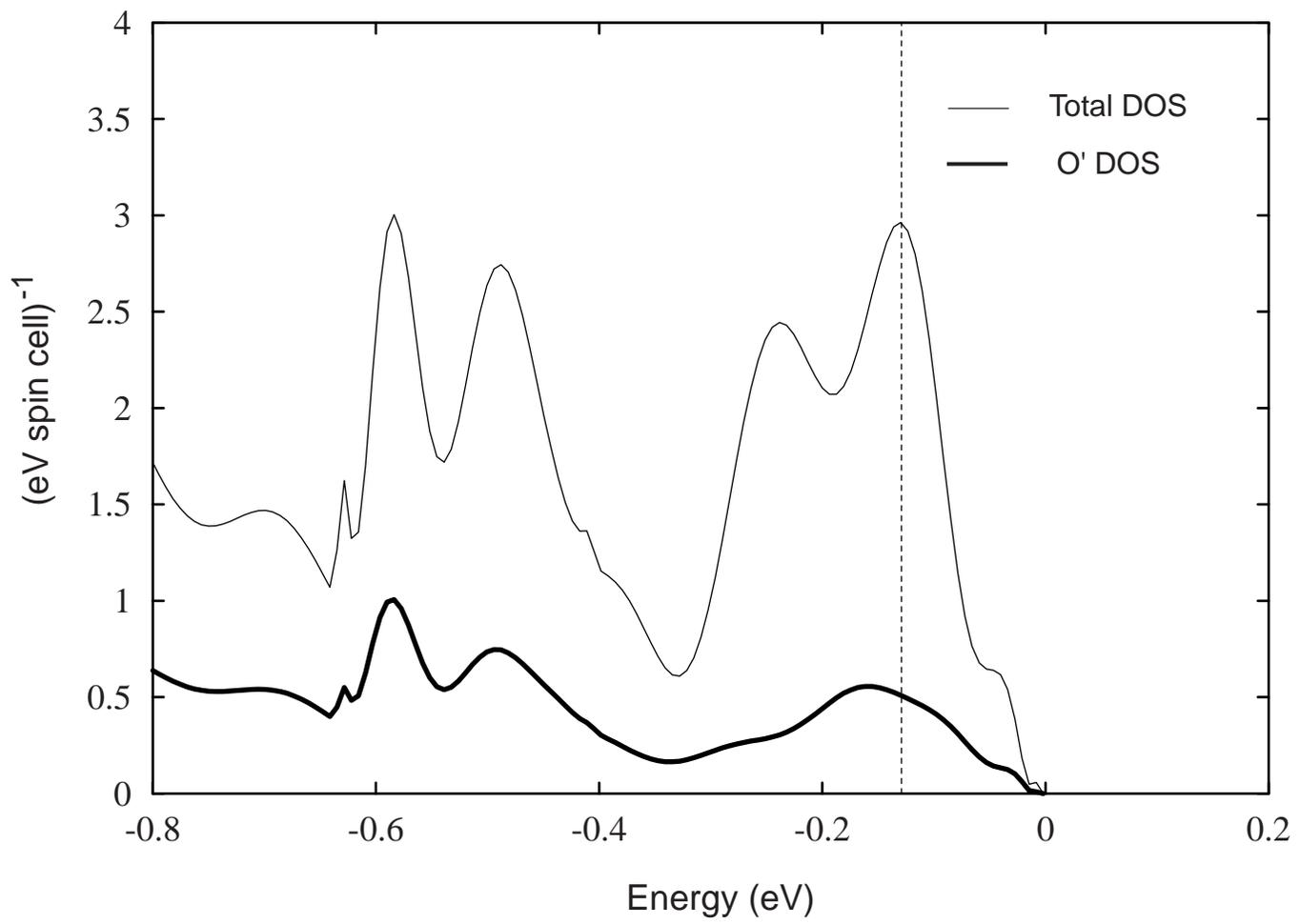
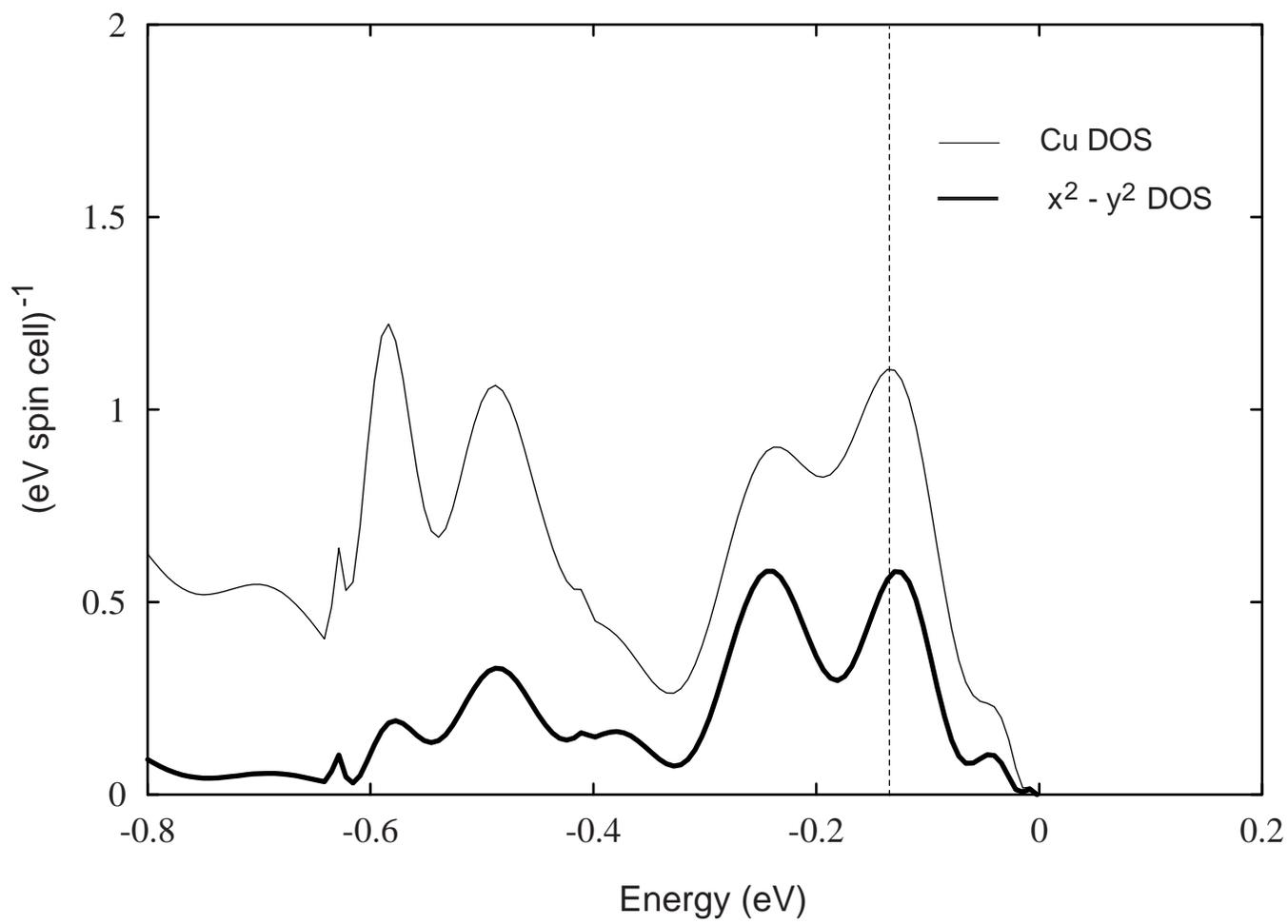
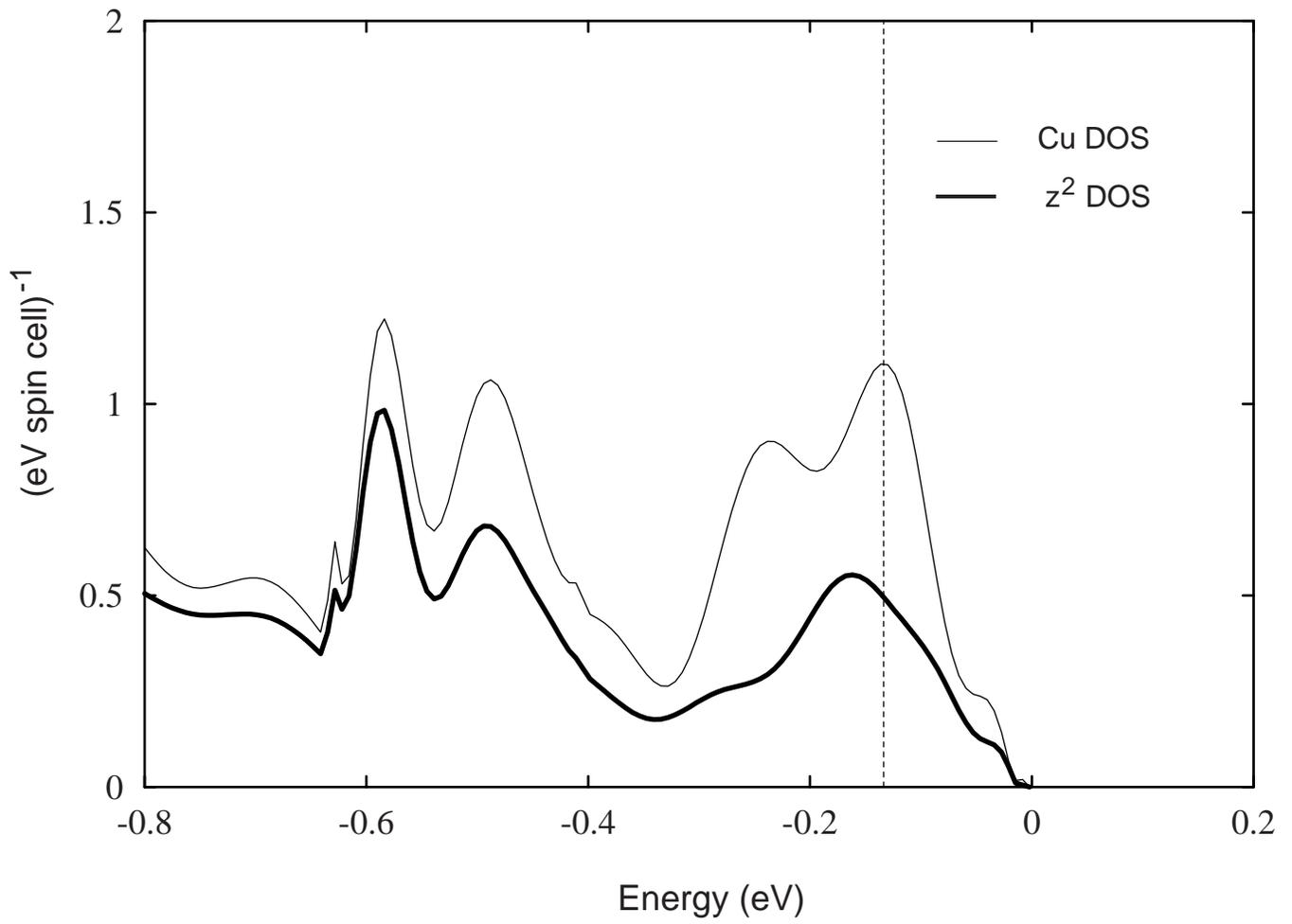
\end